\documentclass[%
reprint,
superscriptaddress,
showpacs,
 amsmath,amssymb,
 aps,
 prl,
]{revtex4-1}

\usepackage{graphicx}% Include figure files
\usepackage{dcolumn}% Align table columns on decimal point
\usepackage{bm}% bold math
\usepackage{amsmath}
\usepackage{autobreak}
\usepackage{hyperref}% add hypertext capabilities
\usepackage[mathlines]{lineno}% Enable numbering of text and display math
\usepackage{natbib}
\usepackage{multirow}
\usepackage{booktabs}

\usepackage{gensymb}

\begin{document}

\preprint{APS/123-QED}

\title{Studies of the Earth shielding effect to direct dark matter searches at the\\ China Jinping Underground Laboratory}% Force line breaks with \\

\affiliation{Key Laboratory of Particle and Radiation Imaging (Ministry of Education) and Department of Engineering Physics, Tsinghua University, Beijing 100084}
\affiliation{Institute of Physics, Academia Sinica, Taipei 11529}

%\affiliation{Department of Physics, Dokuz Eyl\"{u}l University, \.{I}zmir 35160}
\affiliation{Department of Physics, Tsinghua University, Beijing 100084}
\affiliation{NUCTECH Company, Beijing 100084}
\affiliation{YaLong River Hydropower Development Company, Chengdu 610051}
\affiliation{College of Nuclear Science and Technology, Beijing Normal University, Beijing 100875}
\affiliation{College of Physics, Sichuan University, Chengdu 610065}
\affiliation{School of Physics, Peking University, Beijing 100871}
\affiliation{Department of Nuclear Physics, China Institute of Atomic Energy, Beijing 102413}
\affiliation{Sino-French Institute of Nuclear and Technology, Sun Yat-sen University, Zhuhai 519082}
\affiliation{School of Physics, Nankai University, Tianjin 300071}
\affiliation{Department of Physics, Banaras Hindu University, Varanasi 221005}
\affiliation{Department of Physics, Beijing Normal University, Beijing 100875}

\author{Z.~Z.~Liu}
\affiliation{Key Laboratory of Particle and Radiation Imaging (Ministry of Education) and Department of Engineering Physics, Tsinghua University, Beijing 100084}

\author{L.~T.~Yang}\altaffiliation [Corresponding author: ]{yanglt@mail.tsinghua.edu.cn}
\affiliation{Key Laboratory of Particle and Radiation Imaging (Ministry of Education) and Department of Engineering Physics, Tsinghua University, Beijing 100084}
\author{Q. Yue}\altaffiliation [Corresponding author: ]{yueq@mail.tsinghua.edu.cn}
\affiliation{Key Laboratory of Particle and Radiation Imaging (Ministry of Education) and Department of Engineering Physics, Tsinghua University, Beijing 100084}
\author{C.~H.~Yeh}
\altaffiliation{Participating as a member of TEXONO Collaboration}
\affiliation{Institute of Physics, Academia Sinica, Taipei 11529}

\author{K.~J.~Kang}
\affiliation{Key Laboratory of Particle and Radiation Imaging (Ministry of Education) and Department of Engineering Physics, Tsinghua University, Beijing 100084}
\author{Y.~J.~Li}
\affiliation{Key Laboratory of Particle and Radiation Imaging (Ministry of Education) and Department of Engineering Physics, Tsinghua University, Beijing 100084}
\author{M. Agartioglu}
\altaffiliation{Participating as a member of TEXONO Collaboration}
\affiliation{Institute of Physics, Academia Sinica, Taipei 11529}
\author{H.~P.~An}
\affiliation{Key Laboratory of Particle and Radiation Imaging (Ministry of Education) and Department of Engineering Physics, Tsinghua University, Beijing 100084}
\affiliation{Department of Physics, Tsinghua University, Beijing 100084}
\author{J.~P.~Chang}
\affiliation{NUCTECH Company, Beijing 100084}
\author{J.~H.~Chen}
\altaffiliation{Participating as a member of TEXONO Collaboration}
\affiliation{Institute of Physics, Academia Sinica, Taipei 11529}
\author{Y.~H.~Chen}
\affiliation{YaLong River Hydropower Development Company, Chengdu 610051}
\author{J.~P.~Cheng}
\affiliation{Key Laboratory of Particle and Radiation Imaging (Ministry of Education) and Department of Engineering Physics, Tsinghua University, Beijing 100084}
\affiliation{College of Nuclear Science and Technology, Beijing Normal University, Beijing 100875}
\author{W.~H.~Dai}
\affiliation{Key Laboratory of Particle and Radiation Imaging (Ministry of Education) and Department of Engineering Physics, Tsinghua University, Beijing 100084}
\author{Z.~Deng}
\affiliation{Key Laboratory of Particle and Radiation Imaging (Ministry of Education) and Department of Engineering Physics, Tsinghua University, Beijing 100084}

\author{C.~H.~Fang}
\affiliation{College of Physics, Sichuan University, Chengdu 610065}

\author{X.~P.~Geng}
\affiliation{Key Laboratory of Particle and Radiation Imaging (Ministry of Education) and Department of Engineering Physics, Tsinghua University, Beijing 100084}
\author{H.~Gong}
\affiliation{Key Laboratory of Particle and Radiation Imaging (Ministry of Education) and Department of Engineering Physics, Tsinghua University, Beijing 100084}
\author{X.~Y.~Guo}
\affiliation{YaLong River Hydropower Development Company, Chengdu 610051}
\author{Q.~J.~Guo}
\affiliation{School of Physics, Peking University, Beijing 100871}
\author{L. He}
\affiliation{NUCTECH Company, Beijing 100084}
\author{S.~M.~He}
\affiliation{YaLong River Hydropower Development Company, Chengdu 610051}
\author{J.~W.~Hu}
\affiliation{Key Laboratory of Particle and Radiation Imaging (Ministry of Education) and Department of Engineering Physics, Tsinghua University, Beijing 100084}

\author{H.~X.~Huang}
\affiliation{Department of Nuclear Physics, China Institute of Atomic Energy, Beijing 102413}
\author{T.~C.~Huang}
\affiliation{Sino-French Institute of Nuclear and Technology, Sun Yat-sen University, Zhuhai 519082}

\author{H.~T.~Jia}
\affiliation{College of Physics, Sichuan University, Chengdu 610065}
\author{X.~Jiang}
\affiliation{College of Physics, Sichuan University, Chengdu 610065}

\author{H.~B.~Li}
\altaffiliation{Participating as a member of TEXONO Collaboration}
\affiliation{Institute of Physics, Academia Sinica, Taipei 11529}

\author{J.~M.~Li}
\affiliation{Key Laboratory of Particle and Radiation Imaging (Ministry of Education) and Department of Engineering Physics, Tsinghua University, Beijing 100084}
\author{J.~Li}
\affiliation{Key Laboratory of Particle and Radiation Imaging (Ministry of Education) and Department of Engineering Physics, Tsinghua University, Beijing 100084}
%\author{M. X. Li}
%\affiliation{College of Physics, Sichuan University, Chengdu 610065}
%\author{Q. Y. Li}
%\affiliation{College of Physics, Sichuan University, Chengdu 610065}

\author{R.~M.~J.~Li}
\affiliation{College of Physics, Sichuan University, Chengdu 610065}

\author{X.~Q.~Li}
\affiliation{School of Physics, Nankai University, Tianjin 300071}
\author{Y.~L.~Li}
\affiliation{Key Laboratory of Particle and Radiation Imaging (Ministry of Education) and Department of Engineering Physics, Tsinghua University, Beijing 100084}
%\author{Y. F. Liang}
%\affiliation{Key Laboratory of Particle and Radiation Imaging (Ministry of Education) and Department of Engineering Physics, Tsinghua University, Beijing 100084}

\author {B. Liao}
\affiliation{College of Nuclear Science and Technology, Beijing Normal University, Beijing 100875}
\author{F.~K.~Lin}
\altaffiliation{Participating as a member of TEXONO Collaboration}
\affiliation{Institute of Physics, Academia Sinica, Taipei 11529}
\author{S.~T.~Lin}
\affiliation{College of Physics, Sichuan University, Chengdu 610065}
\author{S.~K.~Liu}
\affiliation{College of Physics, Sichuan University, Chengdu 610065}
\author{Y.~Liu}
\affiliation{College of Physics, Sichuan University, Chengdu 610065}
\author {Y.~D.~Liu}
\affiliation{College of Nuclear Science and Technology, Beijing Normal University, Beijing 100875}
\author {Y.~Y.~Liu}
\affiliation{College of Nuclear Science and Technology, Beijing Normal University, Beijing 100875}
\author{H.~Ma}
\affiliation{Key Laboratory of Particle and Radiation Imaging (Ministry of Education) and Department of Engineering Physics, Tsinghua University, Beijing 100084}

\author{Y.~C.~Mao}
\affiliation{School of Physics, Peking University, Beijing 100871}
\author{Q.~Y.~Nie}
\affiliation{Key Laboratory of Particle and Radiation Imaging (Ministry of Education) and Department of Engineering Physics, Tsinghua University, Beijing 100084}
\author{J.~H.~Ning}
\affiliation{YaLong River Hydropower Development Company, Chengdu 610051}
\author{H.~Pan}
\affiliation{NUCTECH Company, Beijing 100084}
\author{N.~C.~Qi}
\affiliation{YaLong River Hydropower Development Company, Chengdu 610051}
\author{J.~Ren}
\affiliation{Department of Nuclear Physics, China Institute of Atomic Energy, Beijing 102413}
\author{X.~C.~Ruan}
\affiliation{Department of Nuclear Physics, China Institute of Atomic Energy, Beijing 102413}

\author{K.~Saraswat}
\altaffiliation{Participating as a member of TEXONO Collaboration}
\affiliation{Institute of Physics, Academia Sinica, Taipei 11529}

\author{V.~Sharma}
\altaffiliation{Participating as a member of TEXONO Collaboration}
\affiliation{Institute of Physics, Academia Sinica, Taipei 11529}
\affiliation{Department of Physics, Banaras Hindu University, Varanasi 221005}
\author{Z.~She}
\affiliation{Key Laboratory of Particle and Radiation Imaging (Ministry of Education) and Department of Engineering Physics, Tsinghua University, Beijing 100084}

\author{M.~K.~Singh}
\altaffiliation{Participating as a member of TEXONO Collaboration}
\affiliation{Institute of Physics, Academia Sinica, Taipei 11529}
\affiliation{Department of Physics, Banaras Hindu University, Varanasi 221005}

\author {T.~X.~Sun}
\affiliation{College of Nuclear Science and Technology, Beijing Normal University, Beijing 100875}

\author{C.~J.~Tang}
\affiliation{College of Physics, Sichuan University, Chengdu 610065}
\author{W.~Y.~Tang}
\affiliation{Key Laboratory of Particle and Radiation Imaging (Ministry of Education) and Department of Engineering Physics, Tsinghua University, Beijing 100084}
\author{Y.~Tian}
\affiliation{Key Laboratory of Particle and Radiation Imaging (Ministry of Education) and Department of Engineering Physics, Tsinghua University, Beijing 100084}

\author {G.~F.~Wang}
\affiliation{College of Nuclear Science and Technology, Beijing Normal University, Beijing 100875}

\author{L.~Wang}
\affiliation{Department of Physics, Beijing Normal University, Beijing 100875}
\author{Q.~Wang}
\affiliation{Key Laboratory of Particle and Radiation Imaging (Ministry of Education) and Department of Engineering Physics, Tsinghua University, Beijing 100084}
\affiliation{Department of Physics, Tsinghua University, Beijing 100084}
\author{Y.~Wang}
\affiliation{Key Laboratory of Particle and Radiation Imaging (Ministry of Education) and Department of Engineering Physics, Tsinghua University, Beijing 100084}
\affiliation{Department of Physics, Tsinghua University, Beijing 100084}
\author{Y.~X.~Wang}
\affiliation{School of Physics, Peking University, Beijing 100871}
\author{Z.~Wang}
\affiliation{College of Physics, Sichuan University, Chengdu 610065}

\author{H.~T.~Wong}
\altaffiliation{Participating as a member of TEXONO Collaboration}
\affiliation{Institute of Physics, Academia Sinica, Taipei 11529}
\author{S.~Y.~Wu}
\affiliation{YaLong River Hydropower Development Company, Chengdu 610051}
\author{Y.~C.~Wu}
\affiliation{Key Laboratory of Particle and Radiation Imaging (Ministry of Education) and Department of Engineering Physics, Tsinghua University, Beijing 100084}
\author{H.~Y.~Xing}
\affiliation{College of Physics, Sichuan University, Chengdu 610065}

%\author{R. Xu}
%\affiliation{Key Laboratory of Particle and Radiation Imaging (Ministry of Education) and Department of Engineering Physics, Tsinghua University, Beijing 100084}

\author{Y.~Xu}
\affiliation{School of Physics, Nankai University, Tianjin 300071}
\author{T.~Xue}
\affiliation{Key Laboratory of Particle and Radiation Imaging (Ministry of Education) and Department of Engineering Physics, Tsinghua University, Beijing 100084}

\author{Y.~L.~Yan}
\affiliation{College of Physics, Sichuan University, Chengdu 610065}

\author{N.~Yi}
\affiliation{Key Laboratory of Particle and Radiation Imaging (Ministry of Education) and Department of Engineering Physics, Tsinghua University, Beijing 100084}
\author{C.~X.~Yu}
\affiliation{School of Physics, Nankai University, Tianjin 300071}
\author{H.~J.~Yu}
\affiliation{NUCTECH Company, Beijing 100084}
\author{J.~F.~Yue}
\affiliation{YaLong River Hydropower Development Company, Chengdu 610051}
\author{M.~Zeng}
\affiliation{Key Laboratory of Particle and Radiation Imaging (Ministry of Education) and Department of Engineering Physics, Tsinghua University, Beijing 100084}
\author{Z.~Zeng}
\affiliation{Key Laboratory of Particle and Radiation Imaging (Ministry of Education) and Department of Engineering Physics, Tsinghua University, Beijing 100084}

\author{B.~T.~Zhang}
\affiliation{Key Laboratory of Particle and Radiation Imaging (Ministry of Education) and Department of Engineering Physics, Tsinghua University, Beijing 100084}
\author {F.~S.~Zhang}
\affiliation{College of Nuclear Science and Technology, Beijing Normal University, Beijing 100875}

%\author{L. Zhang}
%\affiliation{College of Physics, Sichuan University, Chengdu 610065}

\author{Z.~H.~Zhang}
\affiliation{Key Laboratory of Particle and Radiation Imaging (Ministry of Education) and Department of Engineering Physics, Tsinghua University, Beijing 100084}

\author{Z.~Y.~Zhang}
\affiliation{Key Laboratory of Particle and Radiation Imaging (Ministry of Education) and Department of Engineering Physics, Tsinghua University, Beijing 100084}

\author{K.~K.~Zhao}
\affiliation{College of Physics, Sichuan University, Chengdu 610065}
\author{M.~G.~Zhao}
\affiliation{School of Physics, Nankai University, Tianjin 300071}
\author{J.~F.~Zhou}
\affiliation{YaLong River Hydropower Development Company, Chengdu 610051}

\author{Z.~Y.~Zhou}
\affiliation{Department of Nuclear Physics, China Institute of Atomic Energy, Beijing 102413}
\author{J.~J.~Zhu}
\affiliation{College of Physics, Sichuan University, Chengdu 610065}

\collaboration{CDEX Collaboration}
\noaffiliation

\date{\today}% It is always \today, today,

\begin{abstract} Dark matter direct detection experiments mostly operate at deep underground laboratories. It is necessary to consider shielding effect of the Earth, especially for dark matter particles interacting with a large cross section. We analyzed and simulated the Earth shielding effect for dark matter at the China Jinping Underground Laboratory (CJPL) with a simulation package, CJPL Earth Shielding Simulation code (CJPL\_ESS), which is applicable to other underground locations. The further constraints on the $\chi$-N cross section exclusion regions are derived based on the studies with CDEX experiment data.
\begin{description}
\item[PACS numbers]{95.35.+d,
29.40.-n,
98.70.Vc}
\end{description}
\end{abstract}

%\pacs{}% PACS, the Physics and Astronomy
\maketitle

%\tableofcontents

\section{I. Introduction}

Weakly interacting massive particles (WIMPs, denoted as $\chi$) as a dark matter candidate, have not been observed in direct detection experiments, which provided exclusion limits~\cite{cdex102018, darkside, cdmslite, lux, xenon1t}. Direct detection experiments are usually located in deep underground laboratories such as China Jinping Underground Laboratory (CJPL)~\cite{cjpl}, SNOLab~\cite{snolab}, and LNGS~\cite{lngs}, which means that the dark matter particle will go through a few kilometers of rock before reaching the laboratory. Generally, the transport of dark matter is not be affected by rocks, but the shielding effect may not be negligible for a large cross section. Therefore, the flux of dark matter at laboratory may be suppressed when assuming high cross sections. This phenomenon is called Earth shielding effect and has been studied in detail in Refs.~\cite{earthshielding,earthshielding1,earthshielding2,earthshielding3}. 

At large scattering cross section with mean free path in rocks less than a few kilometers, the WIMPs will scatter with nucleus and lose kinetic energy before reaching the detectors at underground locations. Increasing scattering cross sections will reduce the WIMP fluxes above a given velocity. According to Refs.~\cite{earthshielding,earthshielding1,earthshielding2,earthshielding3}, the event rate of WIMP-nucleus ($\chi$-N) scattering will decrease as the cross section increases at large cross section. The direct detection experiments at underground laboratory cannot detect dark matter with a too large $\chi$-N cross section, due to the shielding effect of overburden. That means constraints of $\chi$-N cross section is not an upper limit, but an exclusion region with both upper and lower bounds. References~\cite{earthshielding,earthshielding1,earthshielding2,earthshielding3} have studied the shielding effect in detail, but the geometric model was the entire Earth with the laboratory placed on a spherical shell a few kilometers deep, where the undulating terrain above the laboratory has been ignored. In this way, these works may not be completely suitable for laboratories at traffic tunnel whose overburden are mountains, such as CJPL, LNGS. 
In this paper, we analyze and simulate the Earth shielding effect for dark matter at CJPL, and we also calculated the exclusion region of $\chi$-N cross section with CDEX~\cite{cdex0,cdex1,cdex12014,cdex12016,cdex1b2018,cdex1b_am,cdex102018,cdex10_tech,cdex10_eft} experiment data.

\section{II. Theory}
In the direct search for dark matter, WIMPs are assumed to interact elastically with nucleons, losing a fraction of their kinetic energy and deviating from their original direction. While dark matter particles go through the rocks and scatter multiple times, their tracks are no longer straight lines but broken lines. Consequently, simulations that only have energy loss taken into account are not completely in line with the actual transport process, a complete Monte Carlo (MC) simulation based on $\chi$-N scattering is necessary.

\begin{figure}[!htbp]
\includegraphics[width=\linewidth]{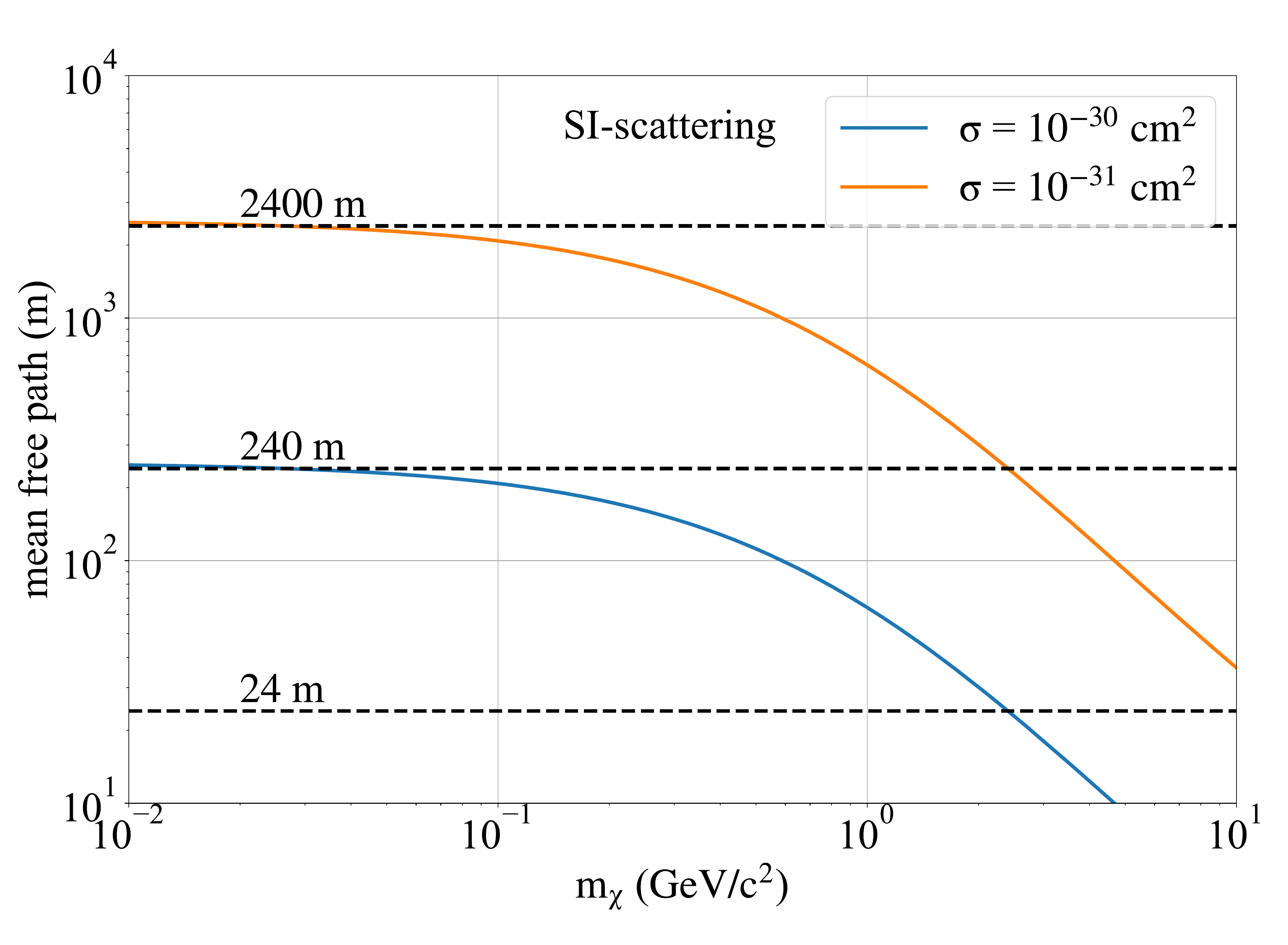}
\caption{
The mean free path of $\chi$ at Jinping Mountain in SI-scattering process, the blue line is that of cross section at $10^{-30}$ $\rm cm^2$, and the orange line is that of cross section at $10^{-31}$ $\rm cm^2$. The density of the transport medium is set to be 2.7 $\rm g/cm^3$, and the composition of the transport medium is adopted as Table \ref{tab::table1}. At the same cross section, the mean free path decreases with the increase of $\chi$'s mass. 
}
\label{fig::mfp}
\end{figure}

Mean free path (denoted as $\lambda$) is a key parameter for MC simulations. The overburden contains multiple nuclides, so the mean free path can be expressed as 
%\begin{small}
\begin{equation}
\begin{aligned}
\frac{1}{\lambda} = \sum_{i} \frac{f_{i} \rho}{m_{A_{i}}} \sigma_{\chi A_{i}},
\end{aligned}
\end{equation}
%\end{small}
where $\rho$ is the density of the transport medium, $f_{i}$ is the mass fraction of the $i_{th}$ nuclide in transport medium, $m_{A_{i}}$ is the mass of the $i_{th}$ nuclide, and $\sigma_{\chi A_{i}}$ is cross section of $\chi$-$A_{i}$ scattering, where $A_{i}$ refers to nucleus of the $i_{th}$ nuclide. For low mass dark matter whose mass $m_{\chi}$ is less than 10 GeV/$c^2$, the form factor $F(q)\sim$1, where $q$ is equal to $\sqrt{2m_{A}E_{R}}$, and $E_R$ is nuclear recoil energy. In this way, $\sigma_{\chi A_{i}}^{\mathrm{SI} }$ is equal to $\sigma_{\chi N}^{\mathrm{SI}}(0) \frac{\mu_{\chi A_i}^2}{\mu_{\chi p}^2} A_{i}^2$ for $\chi$-N spin-independent (SI) scattering, where $\sigma_{\chi N}^{\mathrm{SI}}$ refers to $\chi$-nucleon SI-interactions cross section, and $\mu$ refers to reduced mass. The mean free path of $\chi$ at Jinping Mountain in SI-scattering process was shown in Fig.~\ref{fig::mfp}, where the blue line is that of the cross section at $10^{-30}$ $\rm cm^2$, and the orange line is that of the cross section at $10^{-31}$ $\rm cm^2$. At the same cross section, the mean free path decreases with the increase of $\chi$'s mass. Before every scattering, the free path of dark matter particle can be calculated by
\begin{equation}
\begin{aligned}
l=-\lambda ln(1-\xi),
\end{aligned}
\end{equation}
where $\xi$ follows an uniform distribution between 0 to 1, denoted as $\xi \sim$ U(0,1). 

The $\chi$-N scattering is treated as elastic scattering process. In the center of mass (c.m.) coordinate frame, the magnitude of $\chi$'s velocity remains the same in scattering process, which is expressed as $v_{\chi c}=v_{\chi c}^{'}$, where $v_{\chi c}$ refers to the velocity of $\chi$ in the c.m. frame before scattering, $v_{\chi c}^{'}$ refers to the velocity of $\chi$ in the c.m. frame after scattering. In a laboratory frame, the velocity of $\chi$ after scattering is
\begin{equation}
\begin{aligned}
\vec{v}_{\chi}^{'}=v_{\chi c} \vec{n}+\vec{v}_{c},
\end{aligned}
\end{equation}
where $\vec{v}_{c}$ is the velocity of the c.m. frame in lab frame, $\vec{n}$ is the direction of $\chi$ after scattering in the c.m. frame. The direction of $\chi$ after scattering is isotropic in the c.m. frame, so the scattering angle $\rm cos \alpha \sim U(-1,1)$. The scattering angle $\theta$ in lab frame can be calculated by
\begin{equation}
\begin{aligned}
\mathrm{tan} \theta = \frac{ \mathrm{sin} \alpha}{ \mathrm{cos} \alpha + \frac{m_{\chi}}{m_A}  }.
\end{aligned}
\label{eq::angle}
\end{equation}

\begin{figure}[!htbp]
\includegraphics[width=\linewidth]{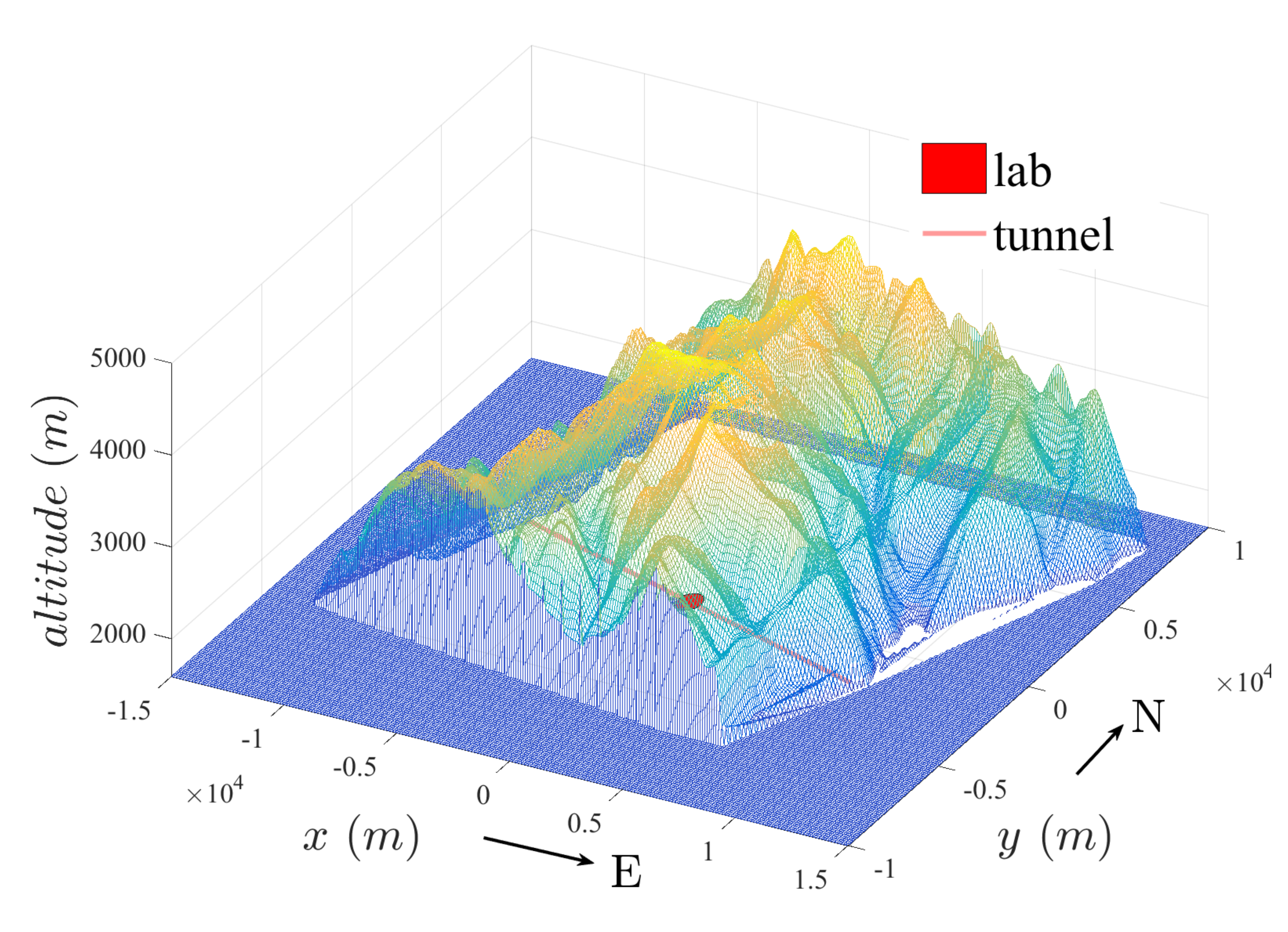}
\caption{
The topographic map of Jinping Mountain, where the red solid line represents the tunnel, the red circular area represents the laboratory. The contour data is obtained from Google Maps. The altitude of the laboratory is about 1600 m, and the highest point of Jinping Mountain is about 4000 m. The coordinate of the laboratory is at (0, 0, 1590 m), and the $x$ axis points to the east, the $y$ axis points to the north, and the $z$ axis points upward. 
}
\label{fig::cjplmountain}
\end{figure}

\begin{figure*}[!htbp]
\includegraphics[width=\linewidth]{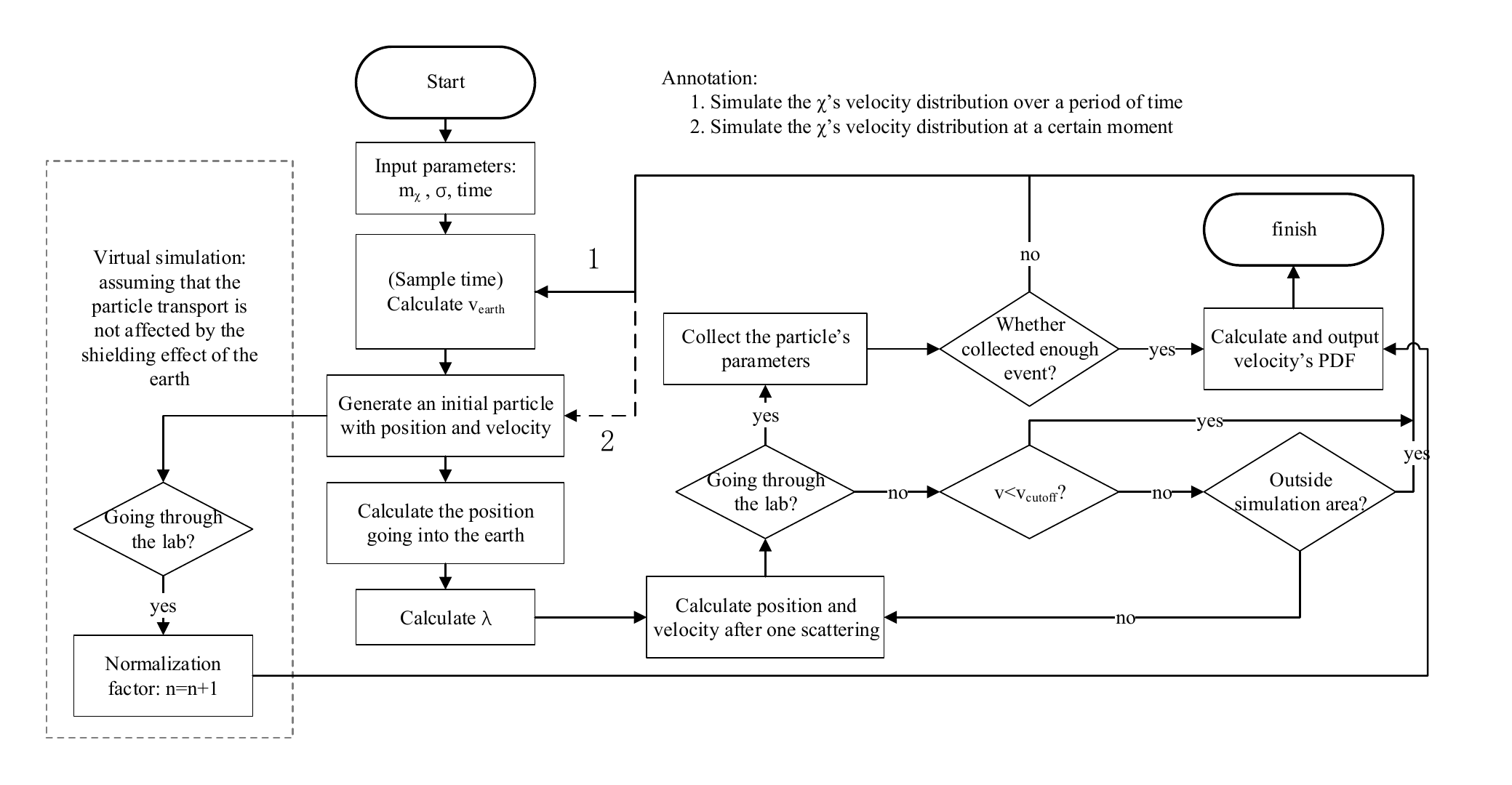}
\caption{
The flow chart of CJPL\_ESS code. The ``virtual simulation'' module serves calculating the number of particles reaching the lab without Earth shielding effect, which can be applied as normalization factor when calculate $\chi$'s velocity distribution. Parallel computing based on the message passing interface and bias sampling method have been used to speed up the program running time. The program has two routes, following route 1 can simulate the $\chi$'s velocity distribution over a period of time, and following route 2 can simulate the $\chi$'s velocity distribution at a certain moment.
}
\label{fig::frame}
\end{figure*}

\section{III. Simulation}

CJPL is located at the center of a traffic tunnel under Jinping Mountain in Sichuan province, southwest China, where the latitude and longitude is approximately ($\rm E101.7\degree$, $\rm N28.2\degree$).  The length of the tunnel is about 17.5 $\rm km$, and the rock overburden at CJPL is about 2400 $\rm m$ vertically. The altitude of the laboratory is about 1600 m, and the highest point of Jinping Mountain is about 4000 m~\cite{cjpl}. The topographic map of Jinping Mountain is shown in Fig.~\ref{fig::cjplmountain}. The rocks of Jinping Mountain are mainly marble, whose density is about 2.7 $\rm g/cm^3$, and the main elements included are O, Ca, Mg, and C. On the basis of the test result of ore composition, the detailed composition of the rocks is shown in Table. \ref{tab::table1}.

\begin{table}[hbp]
\centering
\caption{The main compositions of the rocks of Jinping Mountain, where $f$ represents the mass fraction. The rock is marble, the main component is $\rm CaCO_3$, and the most abundant nuclides are O, Ca, Mg, and C.}
\label{tab::table1}
\begin{tabular*}{\hsize}{@{}@{\extracolsep{\fill}}crrrrrrrrrr@{}}
\toprule
\hline
Element & O & Ca & Mg & C & Si & Al & Fe & K & Na & P \\
\midrule
\hline
$f$ (\%) & 46.42 & 31.96 & 11.50 & 9.59 & 0.19 & 0.15 & 0.10 & 0.07 & 0.01 & 0.01\\
\hline
$Z$ & 8 & 20 & 12 & 6 & 14 & 13 & 26 & 19 & 11 & 15\\
\hline
$A$ & 16 & 40 & 24 & 12 & 28 & 27 & 56 & 39 & 23 & 31\\
\hline
\bottomrule
\end{tabular*}
\end{table}

The Earth model was adopt following Refs.~\cite{earthdensity,earthcomposition}, and the radius ($R_\oplus$) was set to be 6400 km. We spliced the Earth and Jinping Mountain as a geometric model and developed an Earth shielding effect simulation program for CJPL laboratory based on the geometric model, so-called CJPL laboratory Earth shielding effect simulation code (denoted as CJPL\_ESS).

In CJPL\_ESS simulation framework, coordinate system is established with the laboratory stationary and at the origin, as shown in Fig.~\ref{fig::cjplmountain}. The initial dark matter particles are generated on a sphere concentric with the Earth with a radius of $R_0=R_\oplus + h$, where $h$ = 3 km, the center of which defined as point $O_e$. To improve calculation time efficiency, nonuniform sampling is applied to the sampling of initial position $p_0$ of $\chi$. For $\chi$ with given $m_\chi$ and scattering cross section, the $\beta$, which is the angle between $p_0 O_e$ and $y$ axis, was sampled following the probability density distribution function
\begin{equation}
f(x)=\frac{1}{N}e^{ -\frac{R_0}{\lambda_0}(1-x) },
\end{equation}
where $N$ is the normalization factor, $x=\mathrm{cos}\beta$, $\lambda_0$ is the mean free path of $\chi$ in Jinping Mountain. While $\lambda_0\gg R_0$, the initial particles are uniformly generated on the spherical surface. As $\lambda_0$ decreases, the initial particles are more likely to be generated in locations near the laboratory. To ensure that the simulation result is unbiased, the weight of each particle is set to be $1/f(x)$.
The initial velocity is got by 
\begin{equation}
\vec{v}_{\chi , lab}=\vec{v}_{\chi , gal}-\vec{v}_{lab},
\end{equation}
where $\vec{v}_{\chi , gal}$ is the velocity of $\chi$ in the Galaxy, the distribution of which follows Maxwell distribution~\cite{shm}, $\vec{v}_{lab}$ is the velocity of the laboratory in the Galaxy, which is equal to $\vec{v}_{earth}+\vec{v}_{rot}$, sum of the velocity of Earth ($\vec{v}_{earth}$) and the velocity of the Earth’s rotation at the laboratory ($\vec{v}_{rot}$). $\vec{v}_{lab}$ is related to sidereal time and the latitude and longitude of the laboratory, which can be calculated as Ref.~\cite{earthvelocity}. Particles are collected on a horizontal plane with a radius of $r_c$ centered on the laboratory, and the particle that went through the plane are counted to calculate the velocity distribution. 
$r_c$ is set to be 100 km while $\lambda_0 >$ 100 km and set to be 0.5 km, while $\lambda_0 <$ 50 km and decreases from 100 to 0.5 km as $\lambda_0$ from 100 to 50 km.
The flow chart of CJPL\_ESS code is shown in Fig.~\ref{fig::frame}, and the code can output the velocity distribution of $\chi$ at CJPL with a given $m_{\chi}$, $\rm\sigma_{\chi N}$ and time (or a period of time). Parallel computing based on message passing interface and bias sampling method~\cite{biasmethod} have been used to speed up the program running time. 

\begin{figure}[!htbp]
\includegraphics[width=\linewidth]{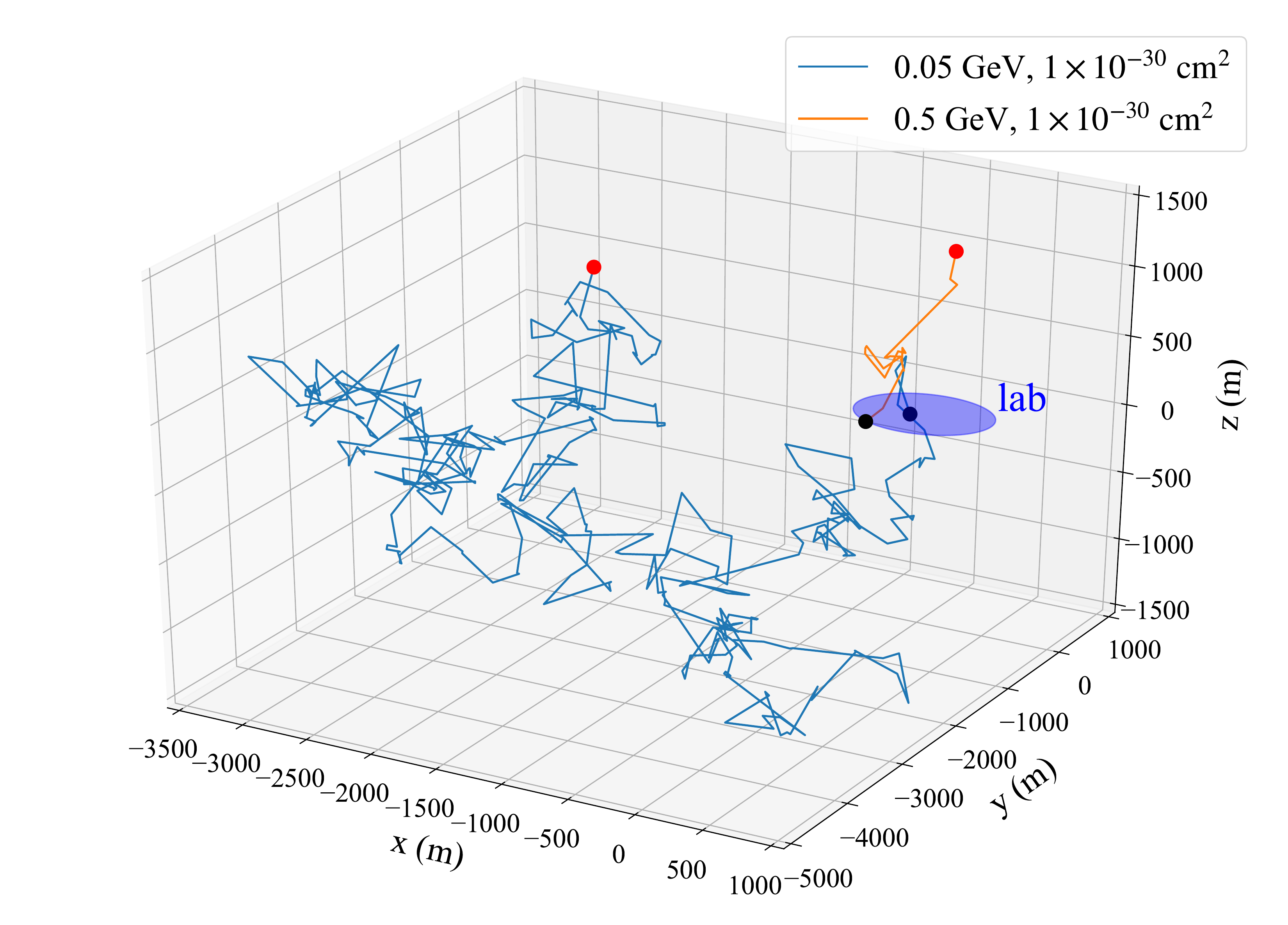}
\caption{
The simulated track (SI scattering) of $\chi$ traveling in the mountain. The location of the laboratory is at the origin. The orange line is the track of $\chi$ at mass of 0.5 GeV and cross section of $1\times10^{-30}$ $\rm cm^2$, and the blue line is the track of $\chi$ at mass of 0.05 GeV and cross section of $1\times10^{-30}$ $\rm cm^2$. According to blue line, light dark matter can reach the lab even after several hundreds of scattering. 
}
\label{fig::track}
\end{figure}

According to Eq.~(\ref{eq::angle}), the scattering angle for WIMP interaction on nucleus can be large, especially for light WIMP. Consequently, the particle will almost completely lose its initial direction after multiple scatterings. The track of two simulated events are shown in Fig.~\ref{fig::track}. The particle at $m_{\chi}$ of 0.05 GeV/$c^2$ and SI cross section of $10^{-30}$ $\rm cm^2$ reached the lab after hundreds of scatterings, that is because the particle can only lose a small part of its total kinetic energy by each scattering, and the maximum rate of energy loss at each scattering is $4\mu_{\chi N}/m_N$. In this way, the direction of particle at lab has a very weak correlation with the initial direction, and the distance traveled is much greater than the displacement. 

\begin{figure}[!htbp]
\includegraphics[width=\linewidth]{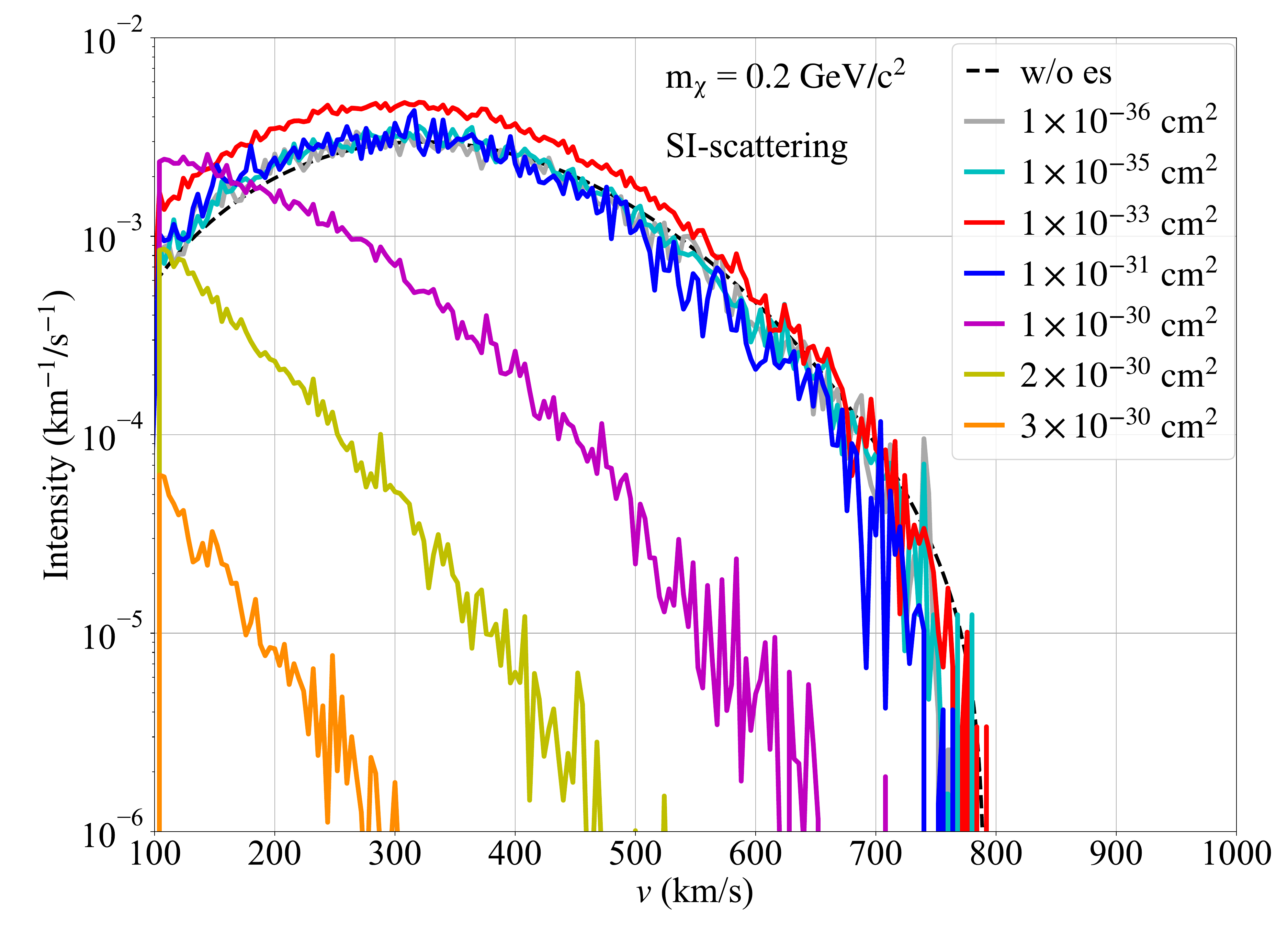}
\caption{
The velocity distribution of $\chi$ at mass of 0.2 GeV in SI-scattering case. The black dashed line is velocity distribution of $\chi$ without Earth shielding effect taken into account, and the solid lines are that at different SI cross section. In simulation, cutoff velocity of particle was set to be 100 km/s.
}
\label{fig::vdis}
\end{figure}

Using CJPL\_ESS code, we simulated the velocity distribution of $\chi$ at different masses and cross section at CJPL. The velocity distributions of $\chi$ normalized to 0.3 GeV/$\rm cm^3$ at mass of 0.2 GeV/$c^2$ and different cross section are shown in Fig. ~\ref{fig::vdis}. It is found that Earth shielding effect becomes noticeable at $\sigma_{\chi \mathrm{N} }^{\mathrm{SI} }>10^{-36}$ $\rm cm^2$. The flux of $\chi$ was slightly enhanced at $\sigma_{\chi \mathrm{N} }^{\mathrm{SI} }$ from $10^{-36}$ $\rm cm^2$ to $10^{-31}$ $\rm cm^2$, because more particles that would not otherwise go through the laboratory after several times scatterings. As $\sigma_{\chi \mathrm{N} }^{\mathrm{SI} }$ increases from $10^{-31}$ $\rm cm^2$, the flux of $\chi$ with velocity greater than the cutoff velocity at lab is getting smaller and smaller.

\section{IV. Result}

CDEX experiment is located at CJPL, and committed to searching dark matter with high purity germanium detector~\cite{cdex0,cdex1,cdex12014,cdex12016,cdex1b2018,cdex1b_am,cdex102018,cdex10_tech,cdex10_eft}. CDEX-1B data and CDEX-10 data are used to Earth shielding effect analysis in this paper. CDEX-1B experiment has for its target a single-element p-type point contact germanium (PPCGe) detector cooled by cold finger, with a fiducial mass of 939 g. A shielding system was built to shield the background radiation from outside the detector, constructed with 20 cm copper, 20 cm PE and 20 cm lead~\cite{cdex1b2018}. Different from CDEX-1B, CDEX-10 experiment used three triple-element PPCGe strings (C10A, B, C) directly immersed in liquid nitrogen (${\rm{LN_2}}$), and the shielding system was constructed with ${\rm{LN_2}}$ and 20 cm copper~\cite{cdex102018}. Compared to that of the overburden, the shielding effect of manually built shielding system to dark matter is negligible, and has been disregarded in simulation process. 

In this work, both $\chi$-N elastic scattering and $\chi$-N inelastic scattering via Migdal effect~\cite{migdaleffect,migdaleffectprl,mec1b} are taken into account. The event rate of $\chi$-N elastic scattering is expressed as
\begin{equation}
\begin{aligned}
\frac {dR}{dE_R} = N_T \frac{\rho_{\chi} }{m_{\chi} } \int dv v f_v(v, \sigma) \frac {d\sigma}{dE_R},
\end{aligned}
\label{eq::nr}
\end{equation}
where $E_R$ is the nuclear recoil energy, $N_T$ is the  number of target nuclei per unit detector mass, $\rho_\chi$ is the density of dark matter, $m_\chi$ is the mass of $\chi$, $v$ is the velocity of $\chi$ in lab frame, $f_v(v,\sigma)$ is velocity distribution of $\chi$, which is the output of CJPL\_ESS code. Considering $\chi$-N SI scattering and spin-dependent (SD) scattering, the differential cross section can be expressed as 
\begin{equation}
\begin{aligned}
\frac {d\sigma}{dE_R}=\frac{m_A}{2 \mu_{\chi A}^2 v^2 } ( \sigma_{\chi A}^{SI} F_{SI}^2(E_R) +  \sigma_{\chi A}^{SD} F_{SD}^2(E_R) ),
\end{aligned}
\end{equation}
where $\mu_{\chi A}$ refers to reduced mass between $\chi$ and nucleus $A$, $\sigma_{\chi A}^{\mathrm{SI} }$ and $\sigma_{\chi A}^{\mathrm{SD} }$ are SI and SD cross section between $\chi$ and $A$, $F$ is the form factor. We will discuss the SI-scattering and SD-scattering case separately later. 

Following Ref. \cite{mec1b}, the event rate of $\chi$-N inelastic scattering via Migdal effect is calculated by 
\begin{small}
\begin{equation}
\begin{aligned}
\frac {dR}{dE_{det}} &= N_T \frac{\rho_{\chi} }{m_{\chi} } \int dv dE_{EM} dE_R\\
 &\times\delta(E_{det}-Q_{nr}E_R-E_{EM}) v f_v(v, \sigma) \frac {d\sigma}{dE_{EM} dE_R},
\end{aligned}
\label{eq::me}
\end{equation}
\end{small}
where $E_{det}$ refers to the energy CDEX PPCGe detectors can detect, which is equal the summation of nuclear recoil energy and electron recoil energy, denoted as $Q_{nr}E_R+E_{EM}$, where $Q_{nr}$ is the quenching factor and was treated following Ref. \cite{mec1b}.

\begin{figure}[!tbp]
\includegraphics[width=\linewidth]{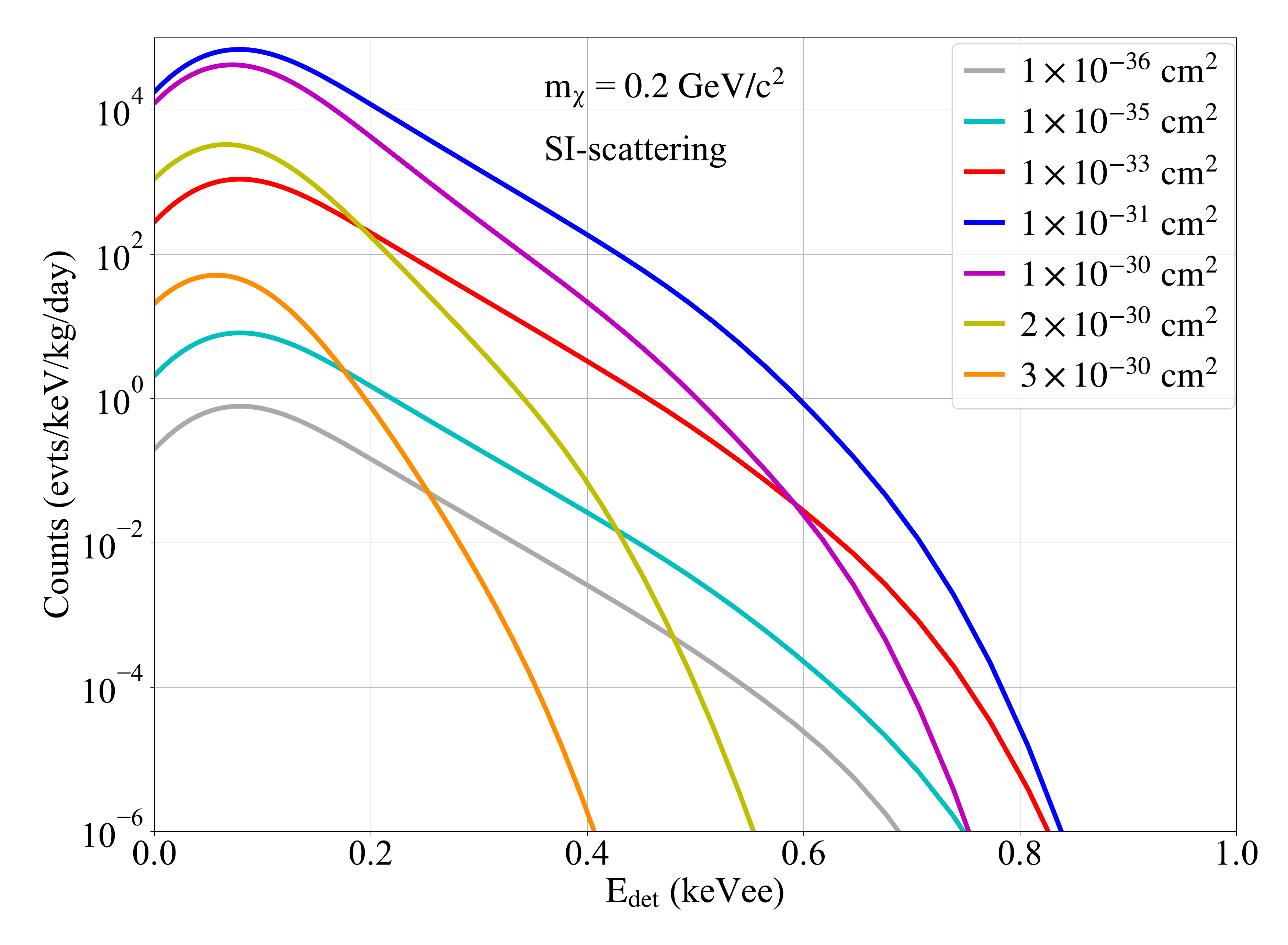}
\caption{
The energy spectrum of $\chi$-N SI inelastic scattering via Migdal effect at $\chi$ mass of 0.2 GeV/$c^2$ in high purity germanium detector, where the SI cross section increases from $1\times 10^{-36}$ to $3\times 10^{-30}$ $\rm cm^2$. The energy resolution is taken into account, the standard deviation of which is 33.5 + 13.2$\times E^{\frac {1}{2}}$ (eV), where $E$ is expressed in keV.
}
\label{fig::spectrum}
\end{figure}

Considering SI scattering, the expected energy spectrum at $\chi$ of different mass and cross section have been calculated using Eqs.~(\ref{eq::nr}) and (\ref{eq::me}). The standard WIMP galactic halo assumption and conventional astrophysical models are used, with $\chi$ density $\rho_\chi$ set to 0.3 GeV/$\rm cm^3$~\cite{shm}. The energy spectrum at mass of 0.2 GeV/$c^2$ with different SI cross sections are shown in Fig. ~\ref{fig::spectrum}. As the SI cross section increasing from $\sim 10^{-36}$ $\rm cm^2$ to $\sim 10^{-30}$ $\rm cm^2$, the count rate of $\chi$-N scattering event increases first and then decreases, and decreases sharply at cross section from $1 \times 10^{-30}$ to $3 \times 10^{-30}$ $\rm cm^2$. 
Migdal effect has been considered in the exclusion line evaluation since it can significantly extend the lower sensitivity reach to light WIMPs at a given detector threshold. Energy loss due to Migdal effects due Earth attenuation is negligible, since the probability is extremely low compared to that for elastic scattering. 
For example, at $m_{\chi}$ = 0.2 GeV/$c^2$, the ratio for mean energy loss of each $\chi$-N scattering via Migdal effect to that of $\chi$-N elastic scattering is $\sim$ 20:1, the ratio of event rate is $\sim$ 1:20000, and the combined ratio of (energy loss $\times$ event rate) is $\sim$ 1:1000.

\begin{figure}[!htbp]
\includegraphics[width=\linewidth]{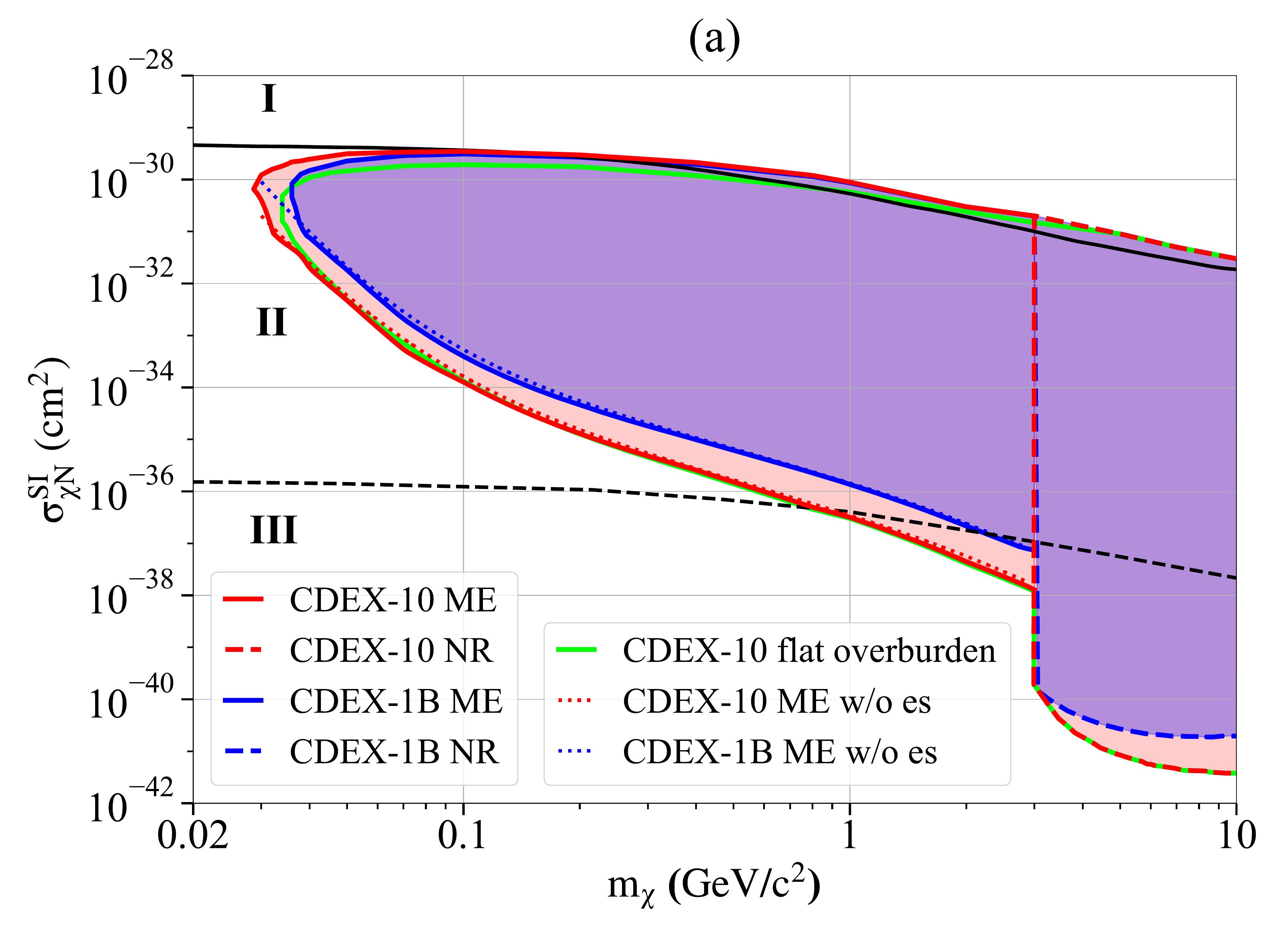}
\includegraphics[width=\linewidth]{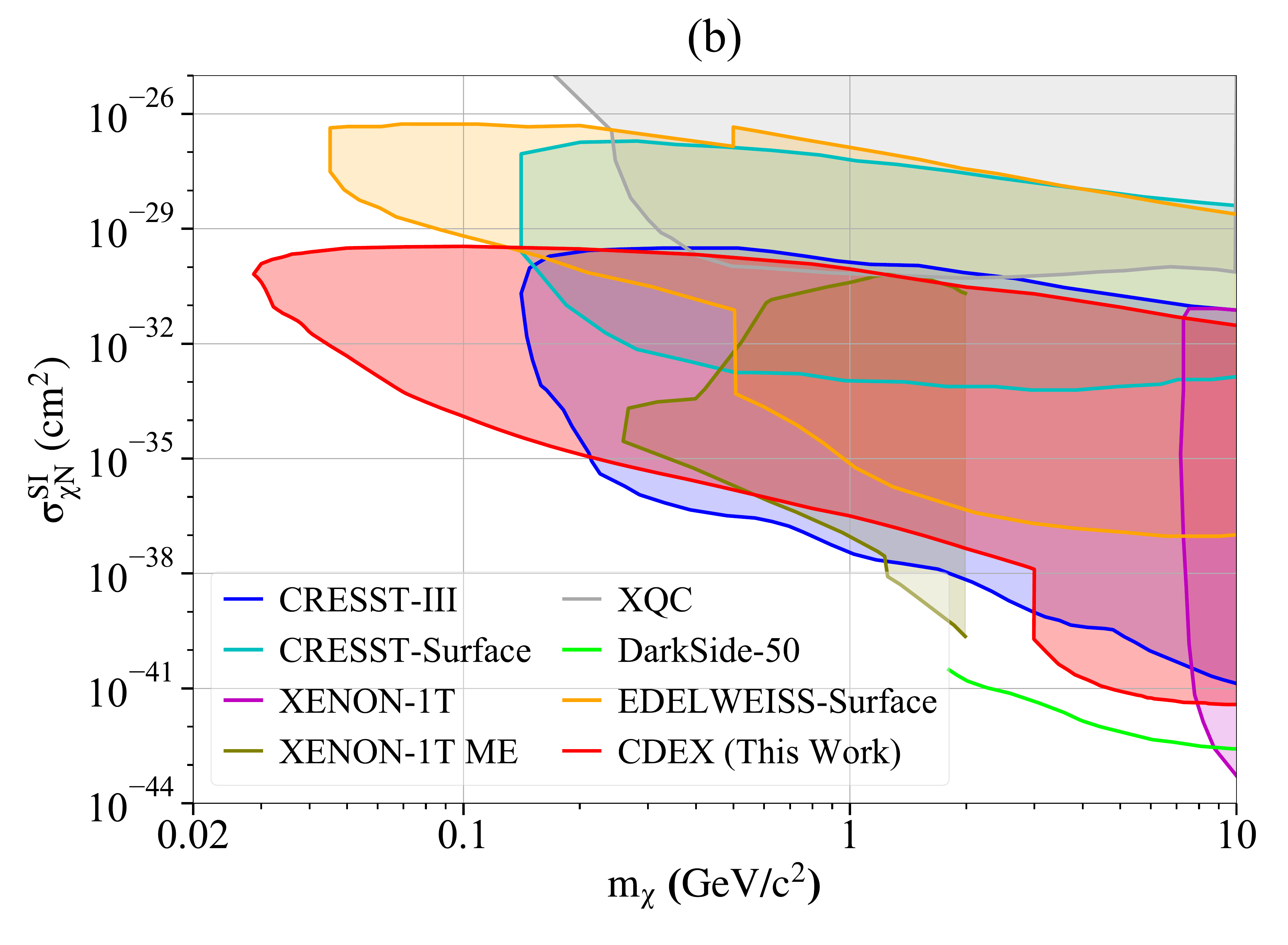}
\caption{
(a) The exclusion region at 90\% CL with Earth shielding effect considered, derived by CDEX data~\cite{cdex1b2018,cdex102018}. The blue region enveloped by blue solid line (``CDEX-1B ME'') is the exclusion region on $\chi$-N scattering via Migdal effect derived by CDEX-1B data, and the red region enveloped by red solid line (``CDEX-10 ME'') is that derived by CDEX-10 data. Corresponding, the regions enveloped by dashed line (``CDEX-1B NR'' and ``CDEX-10 NR'') are the exclusion regions on $\chi$-N elastic scattering derived by CDEX data. The red region and blue region have the same upper limit. The parameter space has been further categorized into three regions by black solid line and black dashed line. As a comparison, the green line is the constraint calculated with flat topography. The dotted lines (``CDEX-1B ME w/o es'' and ``CDEX-10 ME w/o es'') are the constraints without considering Earth shielding effect.  The constraints labeled with "CDEX-1B NR" and "CDEX-1B ME w/o es" are from previous work~\cite{cdex1b2018,mec1b}.
(b)~Summary of results from experiments with published exclusion regions, including CRESST~\cite{earthshielding4}, DarkSide~\cite{darkside}, EDELWEISS~\cite{edelweiss}, XENON-1T~\cite{earthshielding4, xenon1tme}, CDEX, and constraints from the X-ray Quantum Calorimeter experiment (XQC)~\cite{xqc0, xqc}. 
Among these experiments, XENON-1T and CRESST-III are located at LNGS, EDELWEISS-Surface and CRESST-Surface are located at the surface of the Earth, and the Earth shielding effect has been taken into account. 
The upper boundary of ``XENON-1T ME'' is lower than that of CDEX is because Ref.~\cite{xenon1tme} adopted a more conservative calculation method.
}
\label{fig::region}
\end{figure}

The exclusion region at 90\% confidence level (CL) was calculated with CDEX-1B and CDEX-10 data, using binned Poisson method~\cite{binpoisson} and unified approach~\cite{fcmethod}, respectively. The results are shown in Fig.~\ref{fig::region} (a). Although the background and lower limit of the CDEX-1B and CDEX-10 are different, the upper limit of exclusion region of the two experiments are consistent, that's because the count rate of $\chi$-N scattering decreases rapidly as $\sigma_{\chi \mathrm{N} }^{\mathrm{SI} }$ increase to approach the upper limit. In this way, the upper limit of the exclusion region in Fig.~\ref{fig::region} can basically represent the largest cross section for dark matter at the corresponding mass region that can be detected by all experiments at CJPL, within the $\chi$-N SI scattering and context of standard spherical isothermal galactic halo model. In particular, it can be seen in Fig.~\ref{fig::region}(a) that the detailed topography of the Jinping Mountain overburden adopted in this work gives rise to substantial shifts to the upper exclusion bounds (red curve) at the light mass region, relative to those (green curve) from models with flat topography~\cite{earthshielding}. The deviation is $\sim$45\% for WIMP mass at 0.1 GeV/$c^2$. Since it (red curve) is more precise applying detailed topography, we quote it as our official result.

In addition, Fig.~\ref{fig::region} (a) is further categorized into three regions. Region I is where the cross sections are so large that no WIMPs above the cutoff velocity ($<$1\% in flux) would survive. Region III is where all ($>$99\%) WIMPs will arrive laboratory without scattering with the ordinary matter nuclei as they travel through Earth's rock, and region II is between region I and III. Regions I and II are the regions where the Earth shielding effect works. In this way, at $m_\chi <$ 1 GeV/$c^2$, Earth shielding effect should be considered in calculation of exclusion curve.

%%%%%%%%%%%%%%%%SD-analysis

\begin{figure}[!tbp]
\includegraphics[width=\linewidth]{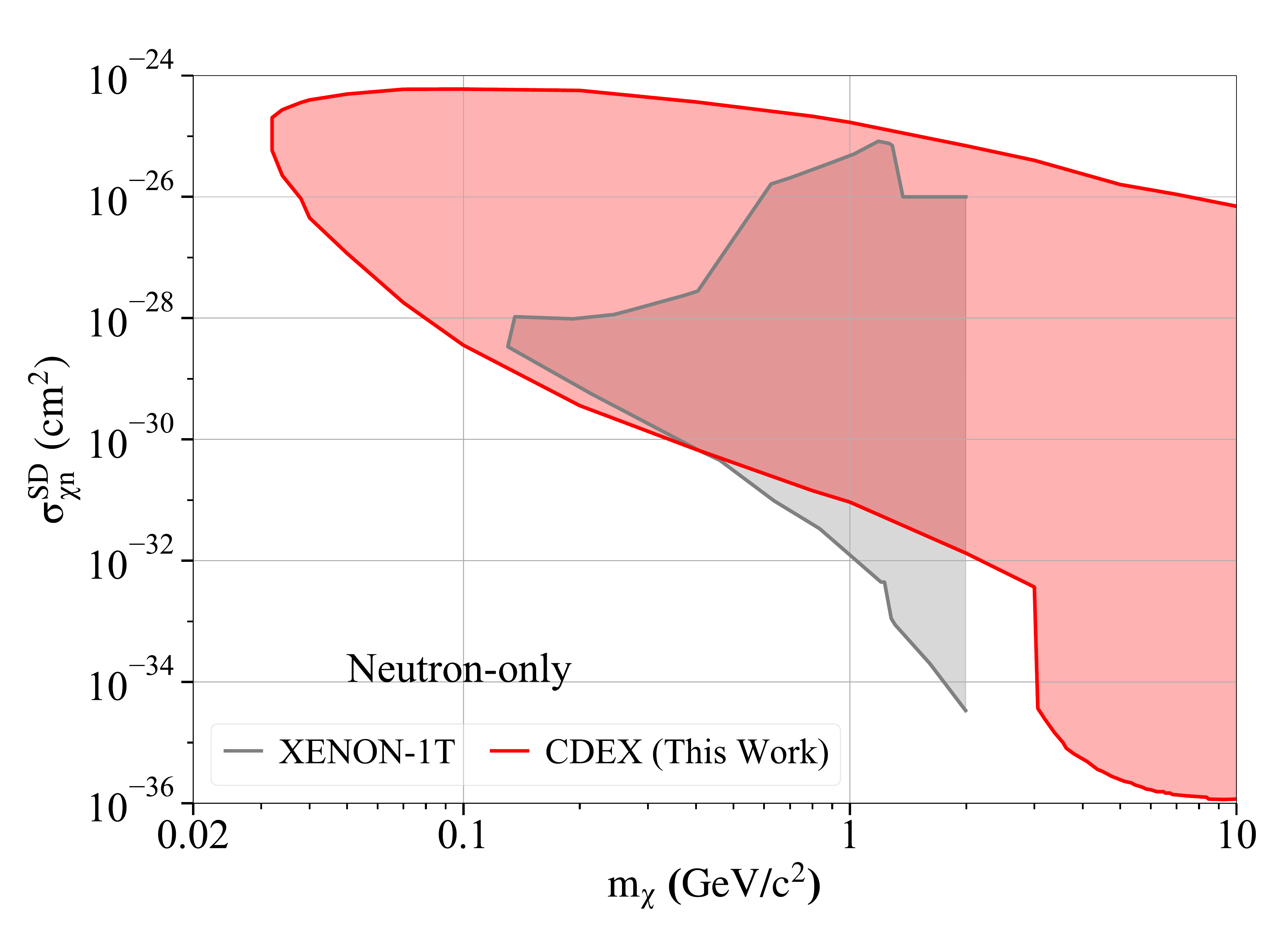}
\caption{
The exclusion region at 90\% CL for $\chi$-N SD scattering of neutron-only case with Earth shielding effect considered, derived by CDEX data~\cite{cdex102018}. The momentum-transfer-dependent spin-structure function of $\rm ^{73}Ge$ was taken from Ref.~\cite{sdparameters_ge73}. The constraint of XENON-1T was taken from Ref.~\cite{xenon1tme}, where a more conservative calculation method has been  adopted in evaluation of Earth shielding effect.
}
\label{fig::regionsd}
\end{figure}

The Earth shielding effect of $\chi$-N SD scattering has also been simulated and evaluated.  
The odd-even nucleus with the largest mass fraction in Jinping Mountain and the Earth have been considered, and the spin expectation values, $\langle S_n \rangle$ and $\langle S_p \rangle$, follow Ref.~\cite{sdformfactor}. The constraints of neutron-only case was calculated and shown in Fig.~\ref{fig::regionsd}. The proton-only case has not been considered, because the expectation value, $\langle S_p \rangle$, of $^{73}\mathrm{Ge}$ approaches zero. The spin expectation values used in SD analysis are shown in Table~\ref{tab::table2}.

\begin{table}[!ht]
\centering
\caption{The spin expectation values used in SD analysis.}
\label{tab::table2}
%\begin{tabular}{clllll}
\begin{tabular*}{\hsize}{@{}@{\extracolsep{\fill}}cclllll@{}}
\toprule
\hline
Element & Location & $Z$ & $A$ & $\langle S_p \rangle$ & $\langle S_n \rangle$ & $J$ \\
\midrule
\hline
C & Jinping & 6 & 13 & -0.009 & -0.172 & 1/2 \\
\hline
Mg & Jinping & 12 & 25 & 0.04 & 0.376 & 5/2 \\
\hline
Al & Jinping & 13 & 27 & 0.343 & 0.0296 & 5/2 \\
\hline
Mg & Mantle & 12 & 25 & 0.04 & 0.376 & 5/2 \\
\hline
Si & Mantle & 14 & 29 & -0.002 & 0.13 & 1/2 \\
\hline
Fe & Core & 26 & 57 & 0 & -0.024 & 1/2 \\
\hline
Ge & Detector & 32 & 73 & 0.030 & 0.378 & 9/2 \\
\hline
\bottomrule
\end{tabular*}
\end{table}

\section{V. Discussion}

In this paper, we have introduced a Monte Carlo simulation package for Earth shielding effect. This package was based on $\chi$-N elastic scattering and developed to be applied to simulate the Earth shielding effect of CJPL (or other underground laboratory that is under a mountain). Compared to previous works~\cite{earthshielding,earthshielding1,earthshielding2,earthshielding3}, the current work includes an effect due to topography of the underground locations. 
This improves on the accuracy of the exclusion bounds. 
To speed up the simulation, nonuniform sampling and bias sampling method have been adopted.
The simulation package is also applicable to other underground locations, after revising the input parameters on topography and compositions. More optional physical processes will be added to this framework, including the inelastic scattering process, DM-electron scattering and so on.

In the framework of CJPL\_ESS, different form factor $F(q)$ can be selected and adopted according to different physical process scenarios. The impact of form factor depends on the energy scale of the momentum transfer during the scattering. In this $\chi$-N SI-scattering and SD-scattering analysis with CDEX data, the approximation of omitting the form factor [$F(q)\sim$1] will bring an error of no more than 1\%. This approximation cannot be applied to $\chi$-N scattering process with large momentum transfer, such as cosmic ray boosted sub-GeV dark matter searches in our subsequent work~\cite{cdexcrdm}, which has a great impact for different ways to handle the form factor.

The studies of Earth shielding effect allow us to derive the upper bounds of the exclusion region on the $\chi$-nucleon scattering cross section, which will make the constraints from underground experiments more complete than the previous results which only presented one exclusion line. In this work, the data of CDEX experiment at CJPL are used in analysis and new constraints on $\chi$-N SI scattering and SD scattering were achieved with the Earth shielding effect taken into account. Further more physics analysis with considering the Earth shielding effect will be carried out, such as modulation effect analysis, cosmic ray boosted sub-GeV dark matter searches~\cite{cdexcrdm} and so on.

\section{ACKNOWLEDGMENTS}

This work was supported by the National Key Research and Development Program of China (Grant No. 2017YFA0402200) and the National Natural Science Foundation of China (Grants No. 12175112, No. 12005111, and No. 11725522). We acknowledge the Center of High performance computing, Tsinghua University for providing the facility support.

\bibliography{espaper.bib}

\end{document}